# Data Stream Clustering: Challenges and Issues

Madjid Khalilian, Norwati Mustapha

*Abstract*— Very large databases are required to store massive amounts of data that are continuously inserted and queried. Analyzing huge data sets and extracting valuable pattern in many applications are interesting for researchers. We can identify two main groups of techniques for huge data bases mining. One group refers to streaming data and applies mining techniques whereas second group attempts to solve this problem directly with efficient algorithms. Recently many researchers have focused on data stream as an efficient strategy against huge data base mining instead of mining on entire data base. The main problem in data stream mining means evolving data is more difficult to detect in this techniques therefore unsupervised methods should be applied. However, clustering techniques can lead us to discover hidden information. In this survey, we try to clarify: first, the different problem definitions related to data stream clustering in general; second, the specific difficulties encountered in this field of research; third, the varying assumptions, heuristics, and intuitions forming the basis of different approaches; and how several prominent solutions tackle different problems.

## I. Introduction

Nowadays we have many applications with massive amount of data which are caused limitation in data storage capacity and processing time. Traditional data mining is not suitable for this kind of applications so they should be tuned and changed or designed with new algorithms. Besides of speed up and storage capacity, real-life concepts tend to change over time:

- Telecommunication and network area: calling records, Network monitoring and traffic engineering, Sensor monitoring & surveillance, Security monitoring, Web logs and Web page click streams
- Business: credit card transaction flows, stock exchange, power supply & manufacturing
- Discovering the evolution of the spread of illnesses. As new cases are reported, finding out how clusters evolve can prove crucial in identifying sources responsible for the spread of illness.
- Discovering the evolution of workload in an e-commerce' server, which can help in dynamically fine tune the server to obtain better performance.
- Discovering meteorological data, such as temperatures registered throughout a region, by observing how clusters of spatial-meteorological points evolve in time.

The growth of volume of existing data and insufficiency of data storage capacity lead us to the dynamic processing data and extracting knowledge. In this way data have been considered as a stream of data which come in from one side and exit from another side so we aren't able to visit data for second time. This main property of data stream arise some difficulties. Two main problems in this area which are related to this property includes: 1) one scan is possible for processing data, 2) data is included evolutionary stream and concepts are changed during the time. It can be gradual or abrupt. Many techniques are used in data mining area but they should be tuned and changed to work in data stream mining. We can categorize data stream mining in three main techniques: classification, clustering and association rules extraction. Many studies have been executed to support data stream mining especially for concept drift[1, 2].

Many researcher's interest is to apply some techniques for increasing compactness of representation, fast and incremental processing of new data points, clear and fast identification of outliers[3]. Scalability and robustness should be studied for data stream mining. Generally it is possible to enumerate two main problems in data stream clustering, concept change and visiting data once.

First of all, How to detect a change in the concepts? Have to:
- Detect the changes as soon as it is occur
- Detect equally well both type of changes abrupt and gradual
- Distinguish between real drift and noise

**What** to **do** if changes are detected?
- **"Forger"** out-of-date examples and clusters (*e.g. Time window*)
- **"Remember"** some of the old clusters and examples

Second problem refers to efficiency. Data stream is similar to river, it means data flow in and flow out. We are unable to visit data twice, so we need to use efficient algorithms.

## II. Gaps

- The architectural aspects of processing data streams have received considerable attention, but most effort are concentrated on the mining and clustering aspects of the problem.
- Algorithms suffer from the ability to handle difficult clustering tasks without supervision. For example, there is no assumption about the number of clusters in data stream but in most methods this parameter should be determined.
- The algorithms required expert assistant in the form of the number of partitions expected or the expected density of clusters.
- They are required to re-learn any recurrently occurring patterns.
- Compactness and separateness of data are the most important problems in the quality of clustering.
- Accuracy in terms of detecting concept drift.
- Efficiency in terms of speed is a vital problem in data mining clustering.
- Previous approaches lack precision in detecting outliers.

M. KHALILIAN is with the Islamic Azad University, Karaj Branch; Iran. He is now PhD candidate in Faculty of Computer Science and Information Technoloy, University Putra Malaysia (UPM). (E-mail: khalilian@ieee.org).

Dr Norwati Mustapha is with the Computer Science Department, Faculty of Computer Science and Information Technoloy, University Putra Malaysia (UPM). (E-mail: Norwati@fsktm.upm.edu.my).

- Uncertain data: in most applications we don't have sufficient data for statistical operations so new methods are needed to manage uncertain data stream in accurate and fast fashion.
- In many applications (i.e. network monitoring), arbitrary shape causes some difficulties in realizing exact clusters of data.
- Data type treatment: different data types (i.e. categorical, ordinal and a mixture of different data types) should be considered in data stream processing.
- Cluster Validity: Recent developments in data stream clustering have heightened the need for determining suitable criteria to validate results. Most outcomes of methods are depended to specific application. However, employing suitable criteria in results evaluation is one of the most important challenges in this arena.
- Space limitation: not only time and concept drift are the main complexity in data stream clustering but also space complexity in some applications (e.g. wireless sensor network monitoring and controlling) can be caused difficulties in processing. Sensors with small memory are not able to keep a big amount of data so new method for data stream clustering should be managed this limitation.
- High dimensional data stream: There are high dimensional data sets (e.g. image processing, personal similarity, customer preferences clustering, network intrusion detection, wireless sensors network and generally time series data) which should be managed through the processing of data stream. In huge databases, data complexity can be increased by number of dimensions.

### III. PRIMITIVE CLUSTERING METHODS

BIRCH can be considered a primitive method in this area [4]. In fact it has been designed for traditional data mining but it is suitable for very large data base so it has been applied for data stream mining. This method introduces two new concepts: micro clustering and macro clustering. Based on these two concepts it could overcome two main difficulties in agglomerative method in clustering: scalability and the inability to undo what was performed in the previous step. It works base on two steps: first it scans data base and builds a tree which is included information about data clusters. In second step BIRCH refines tree by removing sparse nodes as outliers and concrete original clusters. The main disadvantage of this method is the limitation in capacity of leaf. If clusters are not spherical in shape, BIRCH does not perform well because it uses the notion of radius or diameter to control the boundary of a cluster.

STREAM is the next main method which has been designed especially for data stream clustering [5]. In this method K-Medians is leveraged to cluster objects base on SSQ criterion for error measuring. In the first scan objects grouped and medians of each group is gathered and associated them a weight base on the number of objects in the cluster. In next step these medians is clustered until top tree. There are two main disadvantages for this method: time granularity and data evolving.

Recently a novel feature-based method for clustering numeric data streams which employs feature selections has been proposed[6]. When a new point receives in data stream, in absence of each feature, it is assigned to closest cluster and then features are ranked based on point assignments so far in terms of the combination of compactness and separateness quality measures. After that, unimportant features are removed automatically with regard to the ranked list. This method was evaluated in comparison to STREAM. Despite of STREAM it is able to handle high dimensional data. Evolving data stream is not considered like STREAM method.

CluStream addresses these two concerns [7]. In this method ideas in both BIRCH and STREAM are used. Micro clustering and macro clustering are applied in two main components: online component and offline component. It also employs a pyramid structure for organizing macro clusters during the time. Base on this idea it is possible to answer user's question during tilted time. Base on experimental results it has acceptable accuracy and efficiency. In absence of certainty accuracy descend so [8] proposed the UMicro algorithm for clustering uncertain data streams. It has a clear effectiveness with comparison of CluStream Generally approaches which are applied K-Means or K-Medians suffer from lack of accuracy when there are a lot of outliers. Beside, K-Means is also sensitive to value of outliers. These methods are not suitable for discovering clusters with non-convex shapes or clusters of very different size. Number of clusters should be determined as value of parameter K. Moreover, several methods have been proposed to speed up $K$ - means [9, 10] and they can be leveraged for data stream clustering. A spherical $K$ – means SPKMEANS was introduced to address high - dimensional and sparse data objects, particularly for document clustering, which uses a concept vector containing semantic information to represent each cluster [11]. The SPKMEANS algorithm is further cast into the framework of a maximum likelihood estimation of a mixture of $K$ von Mises - Fisher distributions[12]. While $K$ - means has these desirable properties, it also suffers several major drawbacks, particularly the inherent limitations when hill - climbing methods are used for optimization. These disadvantages of $K$ - means attract a great deal of effort from different communities, and as a result, many variants of $K$ -means have appeared to address these obstacles. Finding a globally optimal partition of a given set of documents has been studied and a novel algorithm proposed which is named HKA [13]. In this algorithm harmony search method is utilized for global optimization. The convergence of HKA is studied and proved by using the theory of Markov chain. Modified global K-Means algorithm which is effective for solving clustering problems in gene expression is applied for avoiding local optimal problem [14]. This algorithm computes cluster incrementally and to compute k-partition of data set it uses k-1 cluster centers from the previous iteration. Complexity and computational time is the most important weaknesses of this algorithm.

Fractal Clustering defines clusters as sets of point that exhibit self-similarity [15]. Fractal Clustering (FC) clusters points incrementally, placing them in the cluster in which they have the minimal fractal impact. That is, the cluster that changes the fractal dimension in the least when the point is placed in it. This method can find clusters with arbitrary shapes. It also shows efficient algorithm with managing outlier's concept. Although it is not for data stream clustering but it can be employed in data stream with high dimensions Categorical and ordinal data types. It also doesn't do anything for string or text data.[16] Proposed a new algorithm to handle data stream based on fractal concept. This method is able to group data in arbitrary shape. It also tolerates data with noise and manages high dimensional data. Open issues in this algorithm includes: arbitrary shape at multiple levels of granularity, dynamic adaptation of the parameters in data streams, and investigation of the framework for outlier detection.

## IV. DATA STREAM CLUSTERING METHODS

COBWEB is an incremental clustering technique intends to discover understandable patterns in data [17]. It uses a category function to create a tree. COBWEB keeps a hierarchical clustering model in the form of classification tree. Each node contains a probabilistic description of the concept that summarizes objects classified under that node. Outliers can be managed relatively well in this method but because of the tree structure it includes overhead for managing tree. [18] Proposed a new method base on density clustering to group web pages. They also leveraged F-measure to compute cluster quality. This method is clearly efficient than COBWEB.

A framework base on CluStream to cluster massive text and categorical data has been developed [19]. They compact summary representation of cluster statistics and employ other features which are applied in Clustream. Theirs experimental results show acceptable efficiency.

Density-Based clustering has been leveraged in data stream [20]. This approach can avoid disadvantages of methods base on K-Means. In this method each input data maps into a grid, computes the density of each grid and clusters the grids using a density-based algorithm. This method can discover clusters with arbitrary shapes and detect many outliers around original clusters. It is not able to do anything about categorical and text data.

[21] Extend semantic smoothing model into the text data streams context and present an extended semantic smoothing model. Based on the extended model, two online clustering algorithms OCTS and OCTSM are presented for the clustering of massive text data streams. In these algorithms, they present a new cluster statistics structure named cluster profile, which can capture the real-time semantics of text data Streams and speed up the clustering process at the same time. They also present a series of experimental results illustrating the effectiveness of technique.

[22] Have applied divide and conquer technique in HAC for improving clustering in data stream. It has been mentioned that is not suitable for applications which need speed up in their processing. On the other hand there are no criteria for dividing data and overcoming them. [23] Illustrated a two - level divide - and - conquer clustering algorithm applied to a data set with 2,000 data points. The leader algorithm [24] is first used to form a large number of clusters from the original data. The obtained representatives of these clusters are then clustered with a hierarchical clustering algorithm. [25] Has improved K-Means method by using divide and conquer. Experimental results show that is capable to cluster objects in high quality and efficiency especially in objects with high dimensional.

There is some proposed techniques base on graph theory. Recently [26] have developed a connectivity based reprehensive points to cluster data stream. Accuracy is outstanding in their research but it exhibits low performance. Another point is using a repository for previous data so it is unable to give us a history in different scale time.

[27] Developed a weighted fuzzy CMeans algorithm for data stream. This method is an extension of FCM, it iteratively weighting clusters and data points, and the weighted clusters then incrementally cluster with the next data stream chunk. Effect of outliers has not been considered in weighting clusters.

As it was mentioned, one of the most popular challenges in data stream clustering is outliers detecting. Indeed realizing outliers among evolving data is problematic.[28] Pay more attention to the points detected as outliers and give them a chance of survival in the next incoming chunk of data, rather declare them outlier by observing the current chunk. This method keeps the most suitable candidate outliers. As in data stream we can't keep the entire stream in some physical memory so it ignores the region which is safe and do not contain outliers, and free memory for the next generation of data to be processed effectively.

High dimensionality is one of the major causes in data complexity. Technology makes it possible to automatically obtain a huge amount of measurements. Many studies have been carried out for this purpose. Basically, two main groups of researchers are being focused in this era. One group of researchers has concentrated on static high dimensional datasets [29-33] whereas others pay attention to high dimensional data stream clustering. One of the most popular researches is HPStream [34] that uses the projected clustering approach in order to cluster high dimensional data stream. It selects some features and concrete subspaces to cluster data inside it. Realizing suitable subspace to apply clustering is one major challenge in this algorithm.

## V. CONCLUSIONS

In this paper we have demonstrated some difficulties in data stream clustering. Most data stream applications are high dimensions and new methods need to be developed for this purpose. Furthermore concept drift is nature of data stream and should be managed by new methods. On the other hand data stream is similar to a river: data come in and come out, so we need to design efficient algorithms whereas scan data once and extract hidden patterns inside it. Evolving data, visiting data once and space limitations are major issues in data stream clustering. There are some primitive methods that concreted framework to process data stream but devising new framework with new features other than previous (e.g. using Genetic Algorithm, Ant Colony Algorithm and Neural Networks) can lead us to more effective structures.

## REFERENCES

[1] A. Tsymbal, "The problem of concept drift: definitions and related work," *Informe técnico: TCD-CS-2004-15, Departament of Computer Science Trinity College, Dublin, https://www. cs. tcd. ie/publications/techreports/reports*, vol. 4, pp. 2004-15, 2004.

[2] H. Wang, W. Fan, P. S. Yu, and J. Han, "Mining concept-drifting data streams using ensemble classifiers," 2003.

[3] D. Barbara, "Requirements for clustering data streams," *ACM SIGKDD Explorations Newsletter*, vol. 3, pp. 23-27, 2002.

[4] Zhang, Ramakrishnan, and L. M., "BIRCH: An efficient data clustering method for very large databases " presented at ACM SIGMOD Conference on Management of Data, 1996.

[5] L. O Callaghan, N. Mishra, A. Meyerson, S. Guha, and R. Motwani, "Streaming-data algorithms for high-quality clustering," 2002.

[6] M. J. Asbagh and H. Abolhassani, "Feature-Based Data Stream Clustering," presented at Eigth IEEE/ACIS International Conference on Computer and Information Science, 2009.

[7] C. C. Aggarwal, J. Han, J. Wang, and P. S. Yu, "A framework for clustering evolving data streams," 2003.

[8] C. C. Aggarwal and P. S. Yu, "A framework for clustering uncertain data streams," presented at IEEE 24th International Conference on Data Engineering, Cancún, México, 2008.

[9] D. Pelleg and A. Moore "Accelerating exact K - means algorithms with geometric reasoning " presented at In Proceedings of ACM SIGKDD International Conference on Knowledge Discovery and Data Mining, 1999.

[10] Stoffel and K.Belkoniene, "Parallel K - Means clustering for large data sets ." presented at EuroPar'99 Parallel Processing, Lecture Notes in Computer Science 1685 1999.

[11] I. Dhillon and D. Modha "Concept decompositions for large sparse text data using clustering " *Machine Learning*, vol. 42, pp. 143-175, 2001.

[12] Banerjee and J. Ghosh "Frequency - sensitive competitive learning for scalable balanced clustering on high - dimensional hyperspheres," *IEEE Transactions on Neural Networks* vol. 15, pp. 702-719, 2004.

[13] M. Mahdavi and H. Abolhassani, "Harmony K-means algorithm for document clustering," *Data Mining and Knowledge Discovery*, vol. 18, pp. 370-391, 2009.

[14] A. M. Bagirov and K. Mardaneh, "Modified global k-means algorithm for clustering in gene expression data sets," 2006.

[15] D. Barbara and P. Chen, "Using the fractal dimension to cluster datasets," presented at of the ACM-SIGKDD International Conference on Knowledge and Mining., Boston, 2000.

[16] K. Kailing, H. P. Kriegel, and P. Kroger, "Density-connected subspace clustering for high-dimensional data," 2004.

[17] D. Fisher, "Iterative optimization and simplification of hierarchical clusterings," *Arxiv preprint cs.AI/9604103*, 1996.

[18] M. H. Chehreghani, H. Abolhassani, and M. H. Chehreghani, "Improving density-based methods for hierarchical clustering of web pages," *Data & Knowledge Engineering*, vol. 67, pp. 30-50, 2008.

[19] C. C. Aggarwal and P. S. Yu, "A framework for clustering massive text and categorical data streams," 2006.

[20] Y. Chen and L. Tu, "Density-based clustering for real-time stream data," presented at KDD07, 2007.

[21] X. Li, J. Yang, Q. Wang, J. Fan, and P. Liu, "Research and Application of Improved K-Means Algorithm Based on Fuzzy Feature Selection," 2008.

[22] Guha, Meyerson, A. Mishra, N. Motwani, and O. C. . "Clustering data streams: Theory and practice ." *IEEE Transactions on Knowledge and Data Engineering*, vol. 15, pp. 515-528, 2003.

[23] H. Stahl " Cluster analysis of large data sets," presented at Classification as a Tool of Research , W. Gaul and M. Schader , eds.,, New York, NY : , 1986.

[24] Duda , H. R. , and S. P. , D. , *Pattern classification*, 2 ed: New York,NY : John Wiley & Sons . 2001.

[25] M. Khalilian, N. Mustapha, M. N. Sulaiman, and F. Z. Boroujeni, "K-Means Divide and Conquer Clustering," presented at ICCAE, Thiland, Bangkok, 2009.

[26] S. Lühr and M. Lazarescu, "Incremental clustering of dynamic data streams using connectivity based representative points," *Data & Knowledge Engineering*, vol. 68, pp. 1-27, 2009.

[27] R. Wan, X. Yan, and X. Su, "A weighted fuzzy clustering algorithm for data stream," presented at ISECS International Colloquium on Computing, Communication, Control, and Management.CCCM'08., 2008.

[28] M. Elahi, K. Li, W. Nisar, X. Lv, and H. Wang, "Efficient Clustering-Based Outlier Detection Algorithm for Dynamic Data Stream," presented at Fifth International Conference on Fuzzy Systems and Knowledge Discovery. FSKD'08. , 2008.

[29] C. Fraley and A. E. Raftery, "Model-Based Clustering, Discriminant Analysis, and Density Estimation," *Journal of the American Statistical Association*, vol. 97, pp. 611-632, 2002.

[30] Ali Alijamaat, Madjid Khalilian, and N. Mustapha, "A Novel Approach for High Dimensional Data Clustering," presented at The 3rd International Conference on Knowledge Discovery and Data Mining (WKDD 2010), phuket,Thiland, 2010.

[31] K. Thangavel and A. Pethalakshmi, "Dimensionality reduction based on rough set theory: A review," *Applied Soft Computing Journal*, vol. 9, pp. 1-12, 2009.

[32] J. Peng, C. Tang, D. Yang, J. Zhang, and J. Hu, "Similarity computing model of high dimension data for symptom classification of Chinese traditional medicine," *Applied Soft Computing Journal*, vol. 9, pp. 209-218, 2009.

[33] R. XU and I. DONALD C. WUNSCH, *clustering*: A JOHN WILEY & SONS, INC., PUBLICATION, 2008.

[34] C. C. Aggarwal, J. Han, J. Wang, and P. S. Yu, "A framework for projected clustering of high dimensional data streams," 2004.